\newcommand{\axislim}{1.7}
\newcommand{\outt}{0}
\newcommand{\inn}{0}

\documentclass[journal]{IEEEtran}
\usepackage{amsmath,amssymb,amscd,latexsym,dsfont,psfrag,mathrsfs,tikz}
\usepackage{tabularx,multirow}
\usepackage{graphicx,color,array,cite}
\usepackage{float}
\usepackage{pst-solides3d}
\usepackage{pst-vue3d}
\usepackage{balance}

\newcommand{\bX}{\mathbf{X}}
\newcommand{\bY}{\mathbf{Y}}
\newcommand{\by}{\mathbf{y}}
\newcommand{\bN}{\mathbf{N}}
\newcommand{\E}{\mathbb{E}}
\newcommand{\R}{\mathbb{R}}

\newcommand{\dd}{{\,\mathrm{d}}}

\definecolor{seagreen}{RGB}{46,139,87}

\makeatletter

\newcommand{\Rmnum}[1]{\expandafter\@slowromancap\romannumeral #1@}
\makeatother

\newtheorem{theorem}{Theorem}
\newtheorem{lemma}[theorem]{Lemma}
\newtheorem{corollary}[theorem]{Corollary}

\begin{document}
\title{Achievable Rates of Multidimensional Rotationally Invariant Distributions}

\author{Johnny Karout,~\IEEEmembership{Senior Member,~IEEE,}
Ren\'e-Jean Essiambre,~\IEEEmembership{Fellow,~IEEE,}
Erik Agrell,~\IEEEmembership{Senior Member,~IEEE,}
and Antonia Tulino,~\IEEEmembership{Fellow,~IEEE}

\thanks{
This work was supported by the Swedish Foundation for Strategic Research under Grant RE07-0026,
the Ericsson Research Foundation,
the Friends of Chalmers Foundation, and the
Royal Society of Arts and Sciences in Gothenburg.

J.~Karout was with Bell Labs, Nokia, Holmdel, NJ 07733 USA,
and
Chalmers University of Technology, SE-412 96 Gothenburg, Sweden,
when this work was done. He is now with Ericsson Research, Ericsson AB, 164 80 Stockholm, Sweden (e-mail: johnny.karout@ericsson.com).

E.~Agrell is with the Department of Signals and Systems, Chalmers University of Technology, SE-412 96 Gothenburg, Sweden (e-mail: agrell@chalmers.se).

R.-J.~Essiambre and A.~Tulino are with Bell Labs, Nokia, Holmdel, NJ 07733 USA (e-mail: \{rene.essiambre, a.tulino\}@nokia.com).

}}
\maketitle

\begin{abstract}
The maximum achievable rate or mutual information of multidimensional rotationally invariant distributions in the presence of additive white Gaussian noise is analyzed. A simple expression for the special case of multisphere distributions is derived.
Such distributions consist of points in a multidimensional Euclidean space that are uniformly distributed over several multidimensional concentric hyperspheres.
For the $2$-dimensional case, such distributions have been previously considered in order to reduce the computational complexity of finding a bound on the channel capacity of fiber-optic channels. These distributions take advantage of the statistical rotational invariance of the noise and nonlinear distortions in fiber-optic channels.
Using the derived mutual information expression,
$2$- and $4$-dimensional multisphere distributions are compared for fiber-optic dual-polarization channels dominated by linear noise.
%
%
At high signal-to-noise ratios, $4$-dimensional multisphere distributions offer higher achievable rates than $2$-dimensional ones transmitted on each of the two polarizations of the optical carrier.
Such $4$-dimensional multisphere distributions are also shown to be statistically invariant under $4$-dimensional nonlinear transmission in fibers.

\end{abstract}
\IEEEpeerreviewmaketitle


\section{Introduction}
 \label{sec:intro}
\begin{figure}
\centering \footnotesize
\begin{tikzpicture}[>=stealth,scale=1.1]
\coordinate (O) at (0, 0);

\foreach \point in {O}
   \shadedraw[shading=radial,inner color=black!\outt,outer color=black!\inn,draw=black] (\point) circle (0.2) circle(0.4) circle(0.6) circle(0.8) circle(1.2);

\node[] at (0,-\axislim-0.2) {X polarization};

\node[rotate=45] at (0.7,0.7) {...};
\node[rotate=135] at (-0.7,0.7) {...};
\node[rotate=135+90] at (-0.7,-0.7) {...};
\node[rotate=135] at (0.7,-0.7) {...};

\draw[->,thick] (-\axislim,0) -- (\axislim,0) node[anchor=north ] {$\phi_1(t)$};
\draw[->,thick] (0,-\axislim) -- (0,\axislim) node[anchor=north west] {$\phi_2(t)$};
\end{tikzpicture}
\hfill
\begin{tikzpicture}[>=stealth,scale=1.1]
\coordinate (O) at (0, 0);

\foreach \point in {O}
   \shadedraw[shading=radial,inner color=black!\outt,outer color=black!\inn,draw=black] (\point) circle (0.2) circle(0.4) circle(0.6) circle(0.8) circle(1.2);

\node[] at (0,-\axislim-0.2) {Y polarization};

\node[rotate=45] at (0.7,0.7) {...};
\node[rotate=135] at (-0.7,0.7) {...};
\node[rotate=135+90] at (-0.7,-0.7) {...};
\node[rotate=135] at (0.7,-0.7) {...};

\draw[->,thick] (-\axislim,0) -- (\axislim,0) node[anchor=north ] {$\phi_1(t)$};
\draw[->,thick] (0,-\axislim) -- (0,\axislim) node[anchor=north west] {$\phi_2(t)$};
\end{tikzpicture}
\\[-1ex] (a)

\includegraphics[trim=1cm 2.5cm 0cm 0cm, clip=true,scale=0.5]{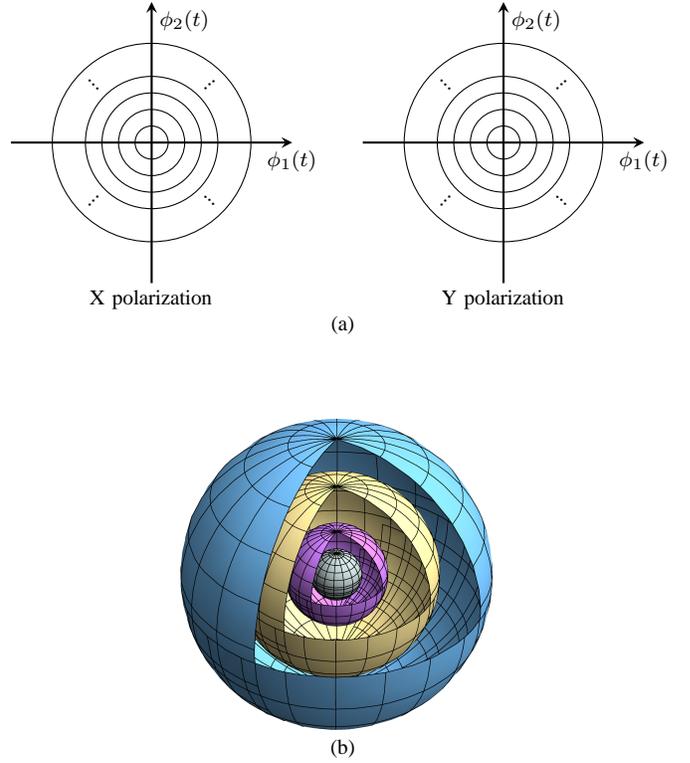}
\\[-3ex] (b)

\caption{(a)~A $2$-D multisphere distribution (multiring distribution) in both polarizations, where $\phi_1(t)$ and $\phi_2(t)$ are the basis functions that span the signal space of conventional in-phase and quadrature modulation formats.
(b)~A $3$-D multisphere distribution, or a cross-section of an $N$-D multisphere distribution.}
\label{fig:ndspheres}

\end{figure}

\IEEEPARstart{T}{he choice} of modulation formats plays a key role in meeting the demands of future communication systems.
Typically, the performance of a modulation format or its equivalent constellation representation in signal space is measured by its spectral efficiency, power efficiency, and complexity. The underlying challenge is to design a constellation that provides the appropriate trade-off between these key performance measures.
%
This draws the attention to the use of multidimensional ($N$-D) constellations, which provide good spectral and power efficiency at the expense of increased complexity.
The dimensions, which can be realized by exploiting diversity in time, frequency (wavelength), space (fiber cores, fiber modes, or antennas), and/or polarizations,
can be looked upon as degrees of freedom used in designing constellations in their respective signal spaces.
For channels where the noise has rotationally invariant statistics,
it is attractive, at least theoretically, to study
input distributions that are also rotationally invariant.
%
The multidimensional Gaussian distribution is the most common such distribution, and it is capacity-achieving over the additive white Gaussian noise (AWGN) channel under an average power constraint \cite[Sec.~24]{Shannon1948}, \cite[Ch.~9]{Cover2006}. Under other circumstances, it may be better to apply input distributions that are discrete in amplitude, but still rotationally invariant, i.e., continuous and uniform in phase (where ``phase'' may be interpreted in a multidimensional sense). Such distributions, which are called \emph{multisphere distributions} or in the $2$-D case \emph{multiring distributions} and exemplified in Fig.~\ref{fig:ndspheres}, have received considerable interest in the literature and are further analyzed in this paper.

Shamai and Bar-David proved in 1995 \cite{Shamai1995} that multiring distributions are capacity-achieving over the $2$-D AWGN channel under simultaneous average and peak power constraints, thereby generalizing a classical $1$-D result by Smith \cite{Smith1971}. Katz and Shamai \cite{Katz2004} proved that multiring distributions achieve the capacity of the partially coherent AWGN channel. Chan \emph{et al.}\ \cite{Chan2005} considered a wide range of channels with peak input constraints and identified several cases where multiring or multisphere distributions are capacity-achieving, including two cases of the Rayleigh-fading channel, the Gaussian interference channel, and a set of parallel AWGN channels with an average total power constraint. Gursoy \emph{et al.} \cite{Gursoy2005} proved similar results for the Rician fading channel under fourth-moment or peak-power input constraints. See references in, e.g., \cite{Katz2004} and \cite{Chan2005} for further examples of discrete-amplitude capacity-achieving distributions.

Another application of multiring distributions arises in fiber-optic communication systems with wavelength-division multiplexing, in which the signals on different wavelengths interfere with each other nonlinearly. It was shown by Ghozlan and Kramer \cite{Ghozlan2010, Ghozlan2011} that by choosing all input distributions as multiring constellations, and under idealized channel conditions, the cross-channel interference can be eliminated.

In~\cite{Freckmann2009, Essiambre2010}, Essiambre \emph{et al.} used multiring distributions to derive lower bounds on the capacity of optical fibers.
%
%
%
%
The analysis is simplified by the fact that all points on the same ring are statistically equivalent.
Therefore, a single probability distribution function can be calculated for each ring by accumulating statistics using all points on that ring. 
The mutual information was numerically computed in~\cite{Freckmann2009} for a nonlinear fiber-optic channel model in an optically-routed network. The study was expanded in~\cite{Essiambre2010}, where results for the AWGN channel were included.
Multiring constellations have also been used to numerically estimate the capacity of the zero-dispersion fiber-optic channel \cite{Yousefi2011} and to lower-bound the capacity of the finite-memory Gaussian noise channel \cite{Agrell2013}.

Coherent optical systems have a $4$-D signal space, corresponding to the two quadratures in both polarizations of an optical signal.
This implies that a $2$-D multisphere distribution can be sent on both polarizations simultaneously, thus increasing spectral efficiency.
A first question that arises is whether a $4$-D multisphere distribution
provides higher achievable rates for the same power efficiency compared to a $2$-D multisphere distribution transmitted on each of the two polarizations.
A second question is whether points on each sphere in a $4$-D multisphere distribution are statistically equivalent for nonlinear propagation in fibers.

In this paper, we analytically derive the mutual information using an $N$-D multisphere distribution for the AWGN channel and display the mutual information results for $2$-D and $4$-D multisphere distributions.
This mutual information serves as a lower bound on the multisphere-constrained capacity, since there is no maximization involved over the radii and the probability of each hypersphere.
Finally, we show the statistical invariance of the stochastic Manakov equations that describe nonlinear propagation over optical fibers under arbitrary $4$-D rotations. This shows that $4$-D hyperspheres are particularly important to consider for fiber-optic communication systems.

\section{System Model and Mutual Information}
 \label{sec:modelandmutualinf}

We consider a discrete-time AWGN channel
\begin{equation}\label{awgn}
\bY=\bX+\bN,
\end{equation}
where $\bX$ is an $N$-D real input vector and $\bN$ is an $N$-D normally distributed noise vector with mean 0 and variance $\sigma^2=N_{0}/2$ per real dimension, where $N_0/2$ is the double-sided power spectral density of the underlying continuous-time noise process.

The mutual information between input and output is~\cite[Sec.~8.4--8.5]{Cover2006}
\begin{equation}\label{eq:mi}
I(\bX;\bY)=h(\bY)-h(\bY|\bX),
\end{equation}
where the differential entropies are
\begin{align}
h(\bY)&=-\int f_{\bY}(\by)  \log_2 f_{\bY}(\by) \dd\by, \label{eq:hy}\\
h(\bY|\bX)&=h(\bN)=\frac{N}{2}\log_2 2 \pi e \sigma^2. \label{eq:hyx}
\end{align}
The maximum mutual information under an average power constraint is the AWGN channel capacity \cite[Eq.~(11)]{Alvarado2015}
\begin{align}\label{cawgn}
C=\frac{N}{2} \log_2 \!\left( 1+\frac{2}{N} A \right),
\end{align}
where $A=\E[\|\bX\|^2]/N_0$ denotes the signal-to-noise ratio (SNR).

\section{Rotationally Invariant Distributions}
 \label{sec:rotinvdist}
We are interested in the case when the input vector $\bX$ is statistically invariant under $N$-D rotations. Hence, its probability density function (pdf) is fully characterized by the pdf of the scalars $\|\bX\|$ or $\|\bX\|^2$. Since the Gaussian vector $\bN$ is also rotationally invariant, so is $\bY$.


The following lemma characterizes the differential entropy of rotationally invariant random variables, generalizing a lemma by Lapidoth and Moser \cite[Eqs.~(319)--(320)]{Lapidoth2003} to arbitrary dimensions $N$.

\begin{lemma}\label{lem:symmetric}
If $\bY$ is a random $N$-D real vector with rotationally invariant statistics, then its differential entropy is%
\footnote{The gamma function satisfies
$$
\Gamma(N/2) =
\begin{cases}
\left(N/2-1\right)!, & \textrm{if $N$ is even,}\\
\frac{\sqrt{\pi}(N-1)!}{2^{N-1}\left(\frac{N-1}{2}\right)!}, & \textrm{if $N$ is odd.}
\end{cases}
$$
}
\begin{align}
h(\bY) &= h(\|\bY\|) + (N-1)\E[\log_2 \|\bY\|] + \log_2\frac{2\pi^{N/2}}{\Gamma(N/2)} \label{eq:hy1}\\
&= h \!\left( \|\bY\|^N \right) + \log_2\frac{\pi^{N/2}}{\Gamma(N/2+1)}. \label{eq:hy2}
\end{align}
\end{lemma}

\begin{IEEEproof}
For any $N$-D real random vector $\bY$, the distribution of its length $R = \|\bY\|$ is calculated by integrating $f_\bY(\by)$ over all $\by$ with the same length, i.e.,
\begin{equation}\label{eq:fydoesntdependony}
f_R(r) = \int_{P_r} f_\bY(\by) \dd\by,
\end{equation}
where $P_r$ is the $N$-D hypersphere with radius $r$ centered at the origin. In the special case when $\bY$ is rotationally invariant, $f_\bY(\by)$ is constant for all points $\by$ with the same radius $\|\by\| = r$. For such distributions, $f_\bY(\by)$ in (\ref{eq:fydoesntdependony}) does not depend on $\by$ and can be moved outside the integral. Thus, for any $\by$,
\begin{equation}\label{eq:frsym}
f_R(\|\by\|) = f_\bY(\by) S_{N-1}(\|\by\|),
\end{equation}
where
\begin{align}\label{eq:spherearea}
S_{N-1}(r) &=\frac{2 \pi^{N/2} r^{N-1}}{\Gamma(N/2)}
\end{align}
is the area of the hypersphere $P_r$~\cite[Sec.~V]{Shannon1959}. 

An integral over $\R^N$ can be written as a double integral: an outer integral over $r$ from 0 to $\infty$ and an inner integral over $P_r$. Therefore, \eqref{eq:hy} can be written as
\begin{align}\label{eq:entropyradiusrepresentation}
h(\bY) &= - \int_0^\infty  \int_{P_r} f_\bY(\by) \log_2 f_\bY(\by) \dd\by \dd r \\
  &=- \int_0^\infty \int_{P_r} \frac{f_R(\|\by\|)}{S_{N-1}(\|\by\|)} \log_2 \frac{f_R(\|\by\|)}{S_{N-1}(\|\by\|)} \dd\by \dd r \label{eq:hy4}\\
  &=- \int_0^\infty \frac{f_R(r)}{S_{N-1}(r)} \log_2 \frac{f_R(r)}{S_{N-1}(r)} \dd r \int_{P_r} \!\!\dd\by \\
  &=- \int_0^\infty f_R(r) \log_2 \frac{f_R(r)}{S_{N-1}(r)} \dd r \label{eq:hy6}\\
  &= h(R) + \E[\log_2 S_{N-1}(R)] \label{eq:hy7},
\end{align}
where \eqref{eq:hy4} follows from \eqref{eq:frsym}.%
\footnote{In~\cite[Eq. (4.12)]{Foschini2003}, an expression similar to \eqref{eq:hy6} was proved for a unit hypersphere. However, the proof above appears to be simpler.} 
Substituting \eqref{eq:spherearea} in \eqref{eq:hy7} completes the proof of \eqref{eq:hy1}.

Let now $T=R^N$. By standard rules for transformation of random variables, $f_R(r) = N r^{N-1} f_T(r^N)$ for all $r\ge 0$. Taking logarithms,
\begin{align}
\log_2 f_R(R) = \log_2 f_T(T) +\log_2 N +(N-1)\log_2 R.
\end{align}
Negating and taking expectations yields
\begin{align}\label{eq:hr}
h(R) = h(T) -\log_2 N - (N-1)\E[\log_2 R].
\end{align}
Substituting \eqref{eq:hr} in \eqref{eq:hy1} and using the gamma function identity $u \Gamma(u) = \Gamma(u+1)$ proves \eqref{eq:hy2}.
\end{IEEEproof}

Using Lemma \ref{lem:symmetric}, an expression for the mutual information between the input and output of the AWGN channel with a given rotationally invariant input distribution can be derived as follows.

\begin{theorem}\label{th:MI-sym}
If $\bX$ has a rotationally invariant distribution, characterized by the distribution $f_{\|\bX\|}$, then the mutual information between $\bX$ and $\bY$ in \eqref{awgn} in bits per $N$-D channel use is
\begin{align} \label{eq:mi-sym}
I(\bX;\bY) &= -\int_{0}^{\infty} f_{\tilde{R}}(\tilde{r}) \log_2 \frac{f_{\tilde{R}}(\tilde{r})}{\tilde{r}^{N-1}} \dd\tilde{r} \nonumber\\
&\qquad + \log_2 \frac{2}{\Gamma(N/2)} - \frac{N}{2} \log_2 2e,
\end{align}
where
\begin{align}
f_{\tilde{R}}(\tilde{r}) &= \int_0^\infty f_{\|\bX\|}(s) \chi\!\left( \tilde{r}, \frac{s}{\sigma} \right) \dd s, \label{eq:r-margin}\\
\chi(\tilde{r},\tilde{s}) &= \frac{\tilde{r}^{N/2}}{\tilde{s}^{N/2-1}} \exp \!\left( - \frac{\tilde{r}^2+\tilde{s}^2}{2} \right) I_{N/2-1} \!\left(\tilde{r} \tilde{s} \right), \label{eq:gdef}
\end{align}
and $I_\nu(u)$ is the $\nu$th order modified Bessel function of the first kind.
\end{theorem}

\begin{IEEEproof}
Substituting \eqref{eq:hyx} and \eqref{eq:hy1} in \eqref{eq:mi} and using $R=\|\bY\|$ yields
\begin{align}
I(\bX;\bY) &= -\int_{0}^{\infty} f_{R}(r) \log_2 \frac{f_{R}(r)}{r^{N-1}} \dd r \nonumber\\
&\qquad + \log_2 \frac{2\pi^{N/2}}{\Gamma(N/2)} - \frac{N}{2} \log_2 2\pi e \sigma^2 \\
&= -\int_{0}^{\infty} f_{R}(r) \log_2 \frac{\sigma^N f_{R}(r)}{r^{N-1}} \dd r \nonumber\\
&\qquad + \log_2 \frac{2}{\Gamma(N/2)} - \frac{N}{2} \log_2 2 e. \label{eq:ixy-sym2}
\end{align}
Defining $\tilde{R} = R/\sigma$, we substitute $f_R(r) = (1/\sigma)f_{\tilde{R}}(r/\sigma)$ and $r=\sigma\tilde{r}$ in \eqref{eq:ixy-sym2} to obtain \eqref{eq:mi-sym}.

From \cite[Eqs.~(1.14), (2.44), and (2.48)]{Simon2002}, the conditional distribution of $Q=\|\bY\|^2$ given $\|\bX\| = s \ge 0$ is the noncentral chi-square distribution with $N$ degrees of freedom
\begin{align}
f_{Q | \|\bX\|} (q|s) &= \frac{1}{2\sigma^2}\left(\frac{q}{s^2}\right)^{N/4-1/2}
  \exp \!\left( - \frac{q+s^2}{2\sigma^2} \right) \nonumber\\
  &\qquad \cdot I_{N/2-1} \!\left( \frac{s\sqrt{q}}{\sigma^2} \right), \quad q\ge 0.
\end{align}
Substituting $Q=\sigma^2 \tilde{R}^2$ yields the noncentral chi distribution
\begin{align} \label{eq:chi}
f_{\tilde{R} | \|\bX\|} (\tilde{r}|s) &= 2\sigma^2 \tilde{r} f_{Q | \|\bX\|} (\sigma^2 \tilde{r}^2|s) \nonumber\\
  &= \chi \!\left(\tilde{r},\frac{s}{\sigma} \right), \quad \tilde{r}\ge 0,
\end{align}
with $\chi$ defined in \eqref{eq:gdef}.
Finally, \eqref{eq:r-margin} follows by $f_{\tilde{R},\|\bX\|}(\tilde{r},s) = f_{\|\bX\|}(s) f_{\tilde{R} | \|\bX\|}(\tilde{r} | s)$ using \eqref{eq:chi}.
\end{IEEEproof}

The following corollaries confine Theorem \ref{th:MI-sym} to the special case of multisphere input distributions. The integral in \eqref{eq:r-margin} is replaced by sums, which reduces the computational complexity.

\begin{corollary}\label{cor:MI-multisphere}
Let $\bX$ be distributed according to an $N$-D multisphere distribution, where the probabilities and radii of each hypersphere $k=1,\ldots,K$ are $p_k$ and $s_k$, respectively. Then the mutual information between $\bX$ and $\bY$ in bits per $N$-D channel use is given by \eqref{eq:mi-sym}, \eqref{eq:gdef}, and
\begin{equation}
f_{\tilde{R}}(\tilde{r}) = \sum_{k=1}^K p_k \chi\!\left( \tilde{r}, \frac{s_k}{\sigma} \right).
\end{equation}
\end{corollary}

\begin{IEEEproof}
Trivial from Theorem \ref{th:MI-sym}.
\end{IEEEproof}

\begin{corollary}\label{cor:MI-uniform}
Let $\bX$ be distributed according to an $N$-D multisphere distribution with $K$
uniformly spaced
hyperspheres
and equal probabilities per hypersphere. Then the mutual information between $\bX$ and $\bY$ in bits per $N$-D channel use as a function of the SNR $A=\E[\|\bX\|^2]/N_0$ is given by \eqref{eq:mi-sym}, \eqref{eq:gdef}, and
\begin{align}
f_{\tilde{R}}(\tilde{r}) &= \frac{1}{K} \sum_{k=1}^K \chi\!\left( \tilde{r}, k \Delta \right), \\
\Delta &= \sqrt{ \frac{12A}{2K^2+3K+1} }. \label{eq:delta}
\end{align}
\end{corollary}

\begin{IEEEproof}
Let $p_k = 1/K$ and $s_k = k \sigma \Delta$ for $k=1,\ldots,K$ in Corollary \ref{cor:MI-multisphere}. The hypersphere spacing $\Delta$ is determined from
\begin{align}
\E[\|\bX\|^2] &= \sum_{k=1}^K p_k s_k^2 \\
  &= \frac{\sigma^2 \Delta^2}{K} \sum_{k=1}^K k^2 \\
  &= \frac{\sigma^2 \Delta^2 \left( 2K^2+3K+1 \right)}{6}. \label{eq:es-uniform}
\end{align}
Equating \eqref{eq:es-uniform} with $\E[\|\bX\|^2] = 2\sigma^2 A$ yields \eqref{eq:delta}.
\end{IEEEproof}

As a further special case, setting $K=1$ and $N=2$ in Corollary \ref{cor:MI-multisphere} or \ref{cor:MI-uniform} gives the constrained capacity of constant-amplitude constellations, i.e., purely phase-modulated transmission schemes, in the presence of AWGN, which was derived by Wyner \cite[Eq.~(12)]{Wyner1966}.

At high SNR, the integrand in \eqref{eq:mi-sym} is very peaky for multisphere input distributions, which can cause numerical problems. These problems can be effectively avoided by sampling these peaks, whose locations are $\tilde{r} \approx \Delta, 2\Delta, \ldots, K \Delta$, denser in the integration algorithm.

\begin{figure}
\centering \footnotesize

\includegraphics[width=\columnwidth]{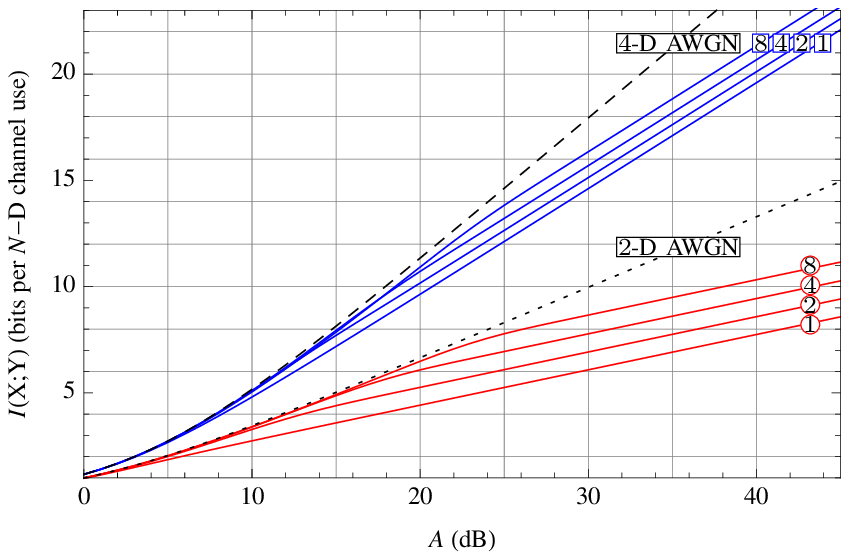}\\[-.5ex] (a)\\[2ex]

\includegraphics[width=\columnwidth]{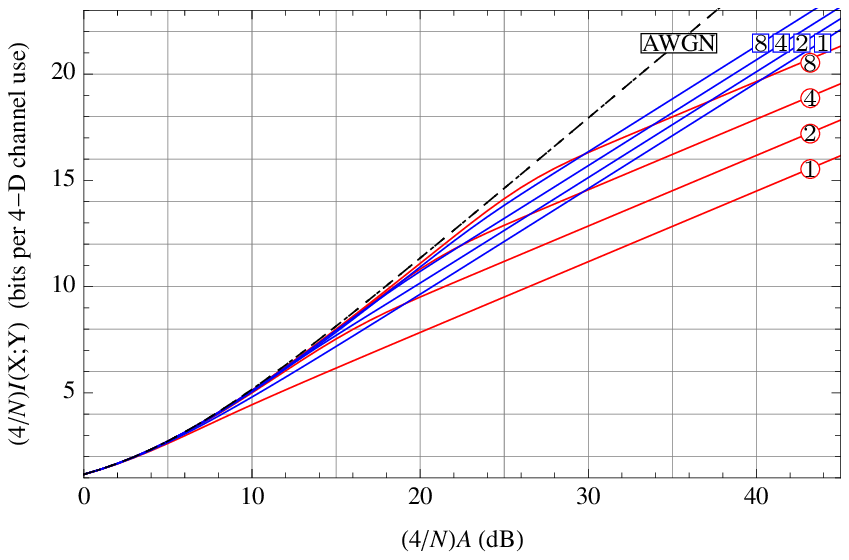}\\[-.5ex] (b)

\caption{(a)~Achievable rates in bits per channel use of $N$-D multisphere distributions, for $N=2$ (circle) and $4$ (rectangle), plotted as functions of SNR $A$.
(b)~The same curves normalized to show rates in bits per $4$-D channel use as functions of $4$-D SNR (i.e., the ratio between energy per four dimensions and $N_0$).}
\label{fig:MI_4d_theoretica}
\end{figure}

\section{Evaluation Using $2$- and $4$-D Multisphere Distributions}
 \label{sec:multisphere2Dand4D}
We evaluate Corollary \ref{cor:MI-uniform} by numerical integration for a distribution of $K$ $N$-D hyperspheres, where $K=\{1,2,4,8\}$, $N=\{2,4\}$, and the probability of transmitting a point belonging to a certain hypersphere is uniform. 
%
%
%
The specific choices of $N$ are motivated by coherent fiber-optic communication with its $4$-D signal space, and noise that can be well approximated by AWGN~\cite{gor63a}.
We compare a $2$-D multisphere distribution transmitted on both polarizations, i.e., using the $4$ available dimensions, and a $4$-D multisphere distribution.

The results are shown in Fig.~\ref{fig:MI_4d_theoretica}, along with the unconstrained capacity \eqref{cawgn} of an $N$-D AWGN channel with $N=2$ and $4$.
At low to medium SNR, the multisphere distributions perform quite close to capacity even with only $2$ or $4$ hyperspheres. The behavior at high SNR, however, differs. It can be observed in Fig.~\ref{fig:MI_4d_theoretica}~(a) that the mutual information of the $4$-D and $2$-D multisphere distribution asymptotically grows as $(3/2)\log_2 A$ and $(1/2)\log_2 A$, respectively, whereas the corresponding capacities \eqref{cawgn} grow as $2\log_2 A$  and $\log_2 A$, respectively. The coefficient of $\log_2 A$, called the \emph{prelog,} is, in analogy with the well-known expression for the multiplexing gain in multiple-input multiple-output communications, equal to the number of complex \emph{degrees of freedom} available for transmission \cite[Chs.~7--9]{Tse2005}. The $N$-D unconstrained AWGN channel \eqref{awgn} has $N$ real degrees of freedom, or equivalently $N/2$ complex degrees of freedom, which explains the observed prelogs $2$ and $1$ for $N=4$ and $2$, respectively.
On the other hand, an $N$-D multisphere distribution has effectively only $N-1$ real degrees of freedom, because the input vector can vary continuously along all $N$ dimensions except the amplitude, which is discrete; this yields a prelog of $(N-1)/2$, which again agrees with the observations in Fig.~\ref{fig:MI_4d_theoretica}~(a).

A perhaps more relevant scenario is to assume that $4$ dimensions are used in all cases, with a fixed amount of energy. The input distribution is either a single $4$-D multisphere distribution or two multiring distributions used in parallel, with half the energy each. Such a comparison is shown in Fig.~\ref{fig:MI_4d_theoretica}~(b), where both axes in Fig.~\ref{fig:MI_4d_theoretica}~(a) have been normalized by a factor $4/N$. In this scenario, the two Gaussian results coincide, since a $4$-D Gaussian distribution is the same as two independent $2$-D Gaussian distributions. At high SNR, $4$-D multisphere distributions are clearly superior to their $2$-D counterparts. This can again be explained by their prelogs, which govern the high-SNR slopes. A single $4$-D multisphere distribution has 3 real degrees of freedom, whereas two $2$-D multiring distributions have only 2 real degrees of freedom together; hence the asymptotic prelogs are $3/2$ and $1$, respectively, whereas the $4$-D Gaussian distribution still has a prelog of $2$. Interestingly, the gain from increasing the number of hyperspheres $K$ is higher if $N=2$ than $4$. This has the somewhat counterintuitive effect that in some cases, for example at $25$ dB and $K=8$, two independent $2$-D multiring distributions have a better performance than a single $4$-D multisphere distribution. The reason is that the combination of two $2$-D distributions yields an amplitude distribution that is more similar to that of a Gaussian $4$-D distribution.

\section{Statistical Rotational Invariance of the Stochastic Manakov Equations}
 \label{sec:Manakov}
In addition to AWGN resulting from optical amplifiers, transmission over optical fibers is
impacted by fiber nonlinearity.
Therefore, it is of great interest to investigate how $4$-D multisphere distributions are impacted by such nonlinearity.
In this section, we show an interesting {\it statistical} invariance of each point on a $4$-D sphere of a multisphere distribution by showing the statistical invariance of the {\it nonlinear} propagation equations of the two complex fields present in optical fibers.

A good approximation for the equations of propagation of the two polarization states supported in a single-mode fiber is a set of coupled partial differential equations referred to as {\it Manakov equations}~\cite{man74}. They are generally written in a vector form~\cite{mar97},
\begin{equation}
    \label{eq:ManakovSMF}
    \frac{\partial\mathbf{E}}{\partial{z}}+ \imath \frac{\beta_2}{2}\frac{\partial^2\mathbf{E}}{\partial{t}^2} - \imath \gamma \frac{8}{9} \|\mathbf{E}\|^2 \mathbf{E} =  \imath \, \mathbf{N},
\end{equation}
where $\mathbf{E}(z,t) = [E_{x}(z,t)~E_{y}(z,t)]^{\rm T}$ with $E_{x}(z,t)$ and $E_{y}(z,t)$ being the two complex fields in the $x$ and $y$ polarizations, respectively,\footnote{We use $\|{\cdot}\|$ in
(\ref{eq:ManakovSMF})
 for consistency with the rest of this paper even though this equation is more commonly written using $|{\cdot}|$ in the physics community.}
as functions of distance $z$ and time $t$,
and where $\mathbf{N}(z,t) = [N_{x}(z,t)~N_{y}(z,t)]^{\rm T}$ with $N_x (z,t)$ and $N_y (z,t)$ being the $x$ and $y$ components of the noise. The noise field $\mathbf{N}(z,t)$ is white in $z$ and $t$ and has a circular Gaussian distribution in each quadrature.

Consider the transformation
\begin{equation}
    \label{eq:unitarytransform}
    \mathbf{E}(z,t) = \mathbf{U} \, \mathbf{E^\prime}(z,t),
\end{equation}
where $\mathbf{U}$ is an arbitrary $2$-D (complex) unitary matrix that is independent of $z$ and $t$. This transformation preserves the norm of $\mathbf{E}(z,t)$, i.e., $\|\mathbf{E^\prime}\| = \|\mathbf{E}\| = \sqrt{E_{x,r}^2 + E_{x,\imath}^2 + E_{y,r}^2 + E_{y,\imath}^2}$, where the indices $r$ and $\imath$ denote the real and imaginary components of the fields, respectively. Preserving the norm means that all points on a hypersphere in the $4$-D space of real fields remain on that hypersphere. Substituting~(\ref{eq:unitarytransform}) in~(\ref{eq:ManakovSMF}) and multiplying the equation by $\mathbf{U}^{-1}$, the new equation expressed in $\mathbf{E^\prime}(z,t)$ assumes the same form as~(\ref{eq:ManakovSMF}) except for the noise term $\mathbf{N}(z,t)$ on the right-hand side, which is now $\mathbf{U}^{-1} \mathbf{N}(z,t)$. One can show that any $2$-D unitary transform operating on a $2$-D complex AWGN field $\mathbf{N}(z,t)$ preserves its statistical properties~\cite[Sec.~7.8.1]{Gallager2008b}. As a result, all $2$-D unitary transformations of the $2$-D complex vector field $\mathbf{E}(z,t)$ leave the Manakov equations~(\ref{eq:ManakovSMF}) statistically invariant. Consequently, all points lying on the same hypersphere of a $4$-D multisphere constellation can be considered as statistically equivalent, even in the presence of fiber nonlinearity. One can exploit this equivalence, for example, to reduce the number of data realizations (i.e. symbols) needed to numerically compute a capacity estimate of a nonlinear system governed by (\ref{eq:ManakovSMF})~\cite{ess13}. One can show that the statistical invariance demonstrated in this section can be readily extended to include additional linear terms in the field $\mathbf{E}(z,t)$ in~(\ref{eq:ManakovSMF}), including higher-order dispersions, fiber loss and amplifier gain. It is interesting to note that a spherical constellation of the form of a multidimensional ball (no discretization of radii) has been studied and shown to provide some level of mitigation of nonlinear effects~\cite{dar2014}.
\section{Conclusions}
 \label{sec:conclusion}
The mutual information of $N$-D rotationally invariant input distributions is derived in the presence of additive white Gaussian noise, by exploiting the rotationally invariant properties of the noise and the distribution itself.
Targeting applications in coherent fiber-optic transmission systems dominated by linear noise, where the signal space is $4$-D due to the two quadratures in both polarizations of the optical carrier, the input distribution is confined to a finite number of concentric, uniformly spaced, equally probable $N$-D hyperspheres.
At high SNR, it is shown that a $4$-D multisphere distribution enables higher achievable rates than a $2$-D multisphere distribution transmitted on both polarizations, even though both distributions make use of the available $4$-D signal space.
Finally,
the statistical invariance of the $4$-D multisphere distribution was shown under the set of partial differential equations describing the nonlinear evolution of the optical fields in optical fibers.

\section{Acknowledgments}
 \label{sec:ack}
We would like to thank G. J. Foschini for his enlightening comments that helped provide a proper direction of this work. We would also like to thank R. W. Tkach for his continuous support.
\balance

\end{document}